# Decentralized Evolution and Consolidation of RDF Graphs

Natanael Arndt and Michael Martin

Agile Knowledge Engineering and Semantic Web (AKSW)
Institute of Computer Science, Leipzig University
Augustusplatz 10, 04109 Leipzig, Germany
{lastname}@informatik.uni-leipzig.de

**Abstract** The World Wide Web and the Semantic Web are designed as a network of distributed services and datasets. In this network and its genesis, collaboration played and still plays a crucial role. But currently we only have central collaboration solutions for RDF data, such as SPARQL endpoints and wiki systems, while decentralized solutions can enable applications for many more use-cases. Inspired by a successful distributed source code management methodology in software engineering a framework to support distributed evolution is proposed. The system is based on Git and provides distributed collaboration on RDF graphs. This paper covers the formal expression of the evolution and consolidation of distributed datasets, the synchronization, as well as other supporting operations.

## 1 Introduction

On the World Wide Web, besides documents, also datasets are getting more and more into the focus. An increasing part of the datasets on the Web is available as Linked Data, also called the *Linked Open Data Cloud*[1] or *Giant Global Graph*[2]. These datasets need to be curated and new data is added in a collaborative way.

Projects from various domains are striving for distributed models to collaborate on common knowledge bases. Examples are in the e-humanities, the *Pfarrerbuch*[3], the *Catalogus Professorum*[4] [25] and the *Heloise – European Network on Digital Academic History*[5] [24]. In libraries meta-data of more and more electronic resources is gathered and shared among stakeholders. The *AMSL*[6] project is looking for collaborative curation and management of electronic resources as

1 http://lod-cloud.net/
2 http://dig.csail.mit.edu/breadcrumbs/node/215
3 http://aksw.org/Projects/Pfarrerbuch
4 http://aksw.org/Projects/CatalogusProfessorum
5 http://heloisenetwork.eu/
6 http://amsl.technology/

Linked Data [2,23]. But even businesses have a need for managing data in distributed setups. In the *LUCID – Linked Value Chain Data*[7] project [13] the communication of data along supply chains is subject of research, and in the *LEDS – Linked Enterprise Data Services*[8] project there is a need for distributed collaboration on datasets to organize business procedures.

Currently available possibilities for collaborating on Linked Data are central SPARQL endpoints and wiki systems [12,11,20]. In both cases a common version of the dataset is kept in a central infrastructure and collaborators are editing on the same computer system simultaneously. This central approach for a synchronized state has drawbacks in different scenarios, where multiple different versions of the dataset are preferable. Multiple different versions of a dataset can exist if a simultaneous access to the central dataset is not possible for all participants for instance due to limited network accessibility. Also it might be, that different levels of access rights are to be implemented, or different releases of a dataset should be available. Even on a common repository it can be intended to have multiple versions of a dataset, for instance in an ongoing discussion where a consensus on a certain topic is not yet reached. Thus it is desirable to manage multiple branches of the dataset evolution.

In the early days of computers and software the term *software crisis* was coined to describe the immaturity of the software engineering process and software engineering domain. Dijkstra described the situation as follows: "*[...] as long as there were no machines, programming was no problem at all; when we had a few weak computers, programming became a mild problem, and now we have gigantic computers, programming had become an equally gigantic problem.*"[9] The process of creating software could be made more reliable and controllable by introducing software engineering methods. In the 1970s software configuration management enabled structured collaborative processes, where version control is an important aspect to organize the evolution of software. Early source code management (SCM), such as *CVS* and *Subversion* allowed to create central repositories. The latest version on the repository represents the current state of development and the linear versioning history draws the evolution process of the software. Decentralized SCM, such as *Darcs*, *Mercurial* and *Git* were developed to allow every member of distributed teams to fork the programs source code and individually contribute new features or bug-fixes as pull-requests which then can be merged into a master branch.

Drawing a parallel from the software engineering to (Linked) Data Engineering, *central* SCM systems would correspond to *central* SPARQL endpoints and wiki systems. To support decentralized and distributed collaboration processes it is necessary to adapt DSCM to Linked Data. This would allow us to support decentralized work, ease the synchronization and manage the independent evolution of datasets. The subject of collaboration in the context of Linked Data are datasets resp. graphs instead of source code files. Similar to source code development with DSCM individual local versions of a dataset are curated by data scientists and domain experts. For consolidating those diverged datasets, a synchronization resp. merging strategy needs to be applied. Berners-Lee and Connolly also deal with *the synchronization problem* [6] and plead for not mixing the individual parts of the problem, to perhaps get a more robust and extensible system. On the informal *Semantic Web Layer Cake* model[10] we can locate different layers, where a certain level of consistency has to be considered: *data interchange*; *ontology*, *rule* and *RDFS* (*ontological*); and *applications*. Considering the *data interchange* layer it is important, that valid RDF data is always provided. This RDF data can then serve as input for checking the compliance to *ontological* rules. Additionally, requirements to the data might be assumed on the *application* layer.

[7] `http://www.lucid-project.org/`

[8] `http://www.leds-projekt.de/`

[9] `https://www.cs.utexas.edu/users/EWD/transcriptions/EWD03xx/EWD340.html`

In the ecosystem around DSCM continuous integration systems are responsible for verifying each operation on the source code, e.g. commits of a pull-request or the result of a merge process. These systems execute among others unit and integration tests before they send a positive or negative result telling the maintainers whether to accept or reject changes to the repository. For managing Semantic Web data continuous integration tools can be responsible for verifying the resulting graph on the *ontological* and *application* layer, as presented in [22] and [19].

In this paper we concentrate on the syntactic *data interchange* layer of the problem and present the following contributions. We propose a formalized model for expressing change and evolution with support for tracking, reverting, branching and merging changes. This model is generic, quad aware and supports blank nodes. To actually pursue merging operations, identifying and resolving conflicts we propose various merge strategies, which can be utilized in different scenarios. Moreover, we evaluate our model regarding correctness and performance using a respective *Git* based implementation. The paper is structured as follows. The state of the art and related work are presented and discussed in section 2. An introduction to the preliminaries for the formal model with basic definitions is given in section 3. The operations on a version graph of distributed evolving datasets are defined in more detail in section 4. Different merge strategies are presented in section 5. The presented concepts are evaluated regarding correctness and performance, using our prototypical implementation, in section 6. Finally a conclusion is given together with a prospect to future work in section 7.

## 2 Related Work

First we consider abstract models for expressing changes and evolution, and second, we examine implementations dealing with versioning of RDF data. Berners-Lee and Connolly [6] give a general overview on the problem of synchronization and on how to calculate delta on RDF graphs. This work concentrates on possibilities to transfer changes to datasets by applying patches. They introduce an ontology that describes patches in *"a way to uniquely identify what is changing"* and *"to distinguish between the pieces added and those subtracted"*. Haase and Stojanovic [16] introduce their concept of ontology evolution as follows: *"Ontology evolution can be defined as the timely adaptation of an ontology to the arisen changes and the consistent management of these changes. [...] An important aspect in the evolution process is to guarantee the consistency of the ontology when changes occur, considering the semantics of the ontology change."*. This paper concentrates on the linear evolution process of an individual dataset, rather than the distributed evolution process. For dealing with inconsistency resp. consistency they define three levels: structural, logical, and user-defined consistency. In the remainder of the paper they mainly concentrate on the implications of the evolution with respect to OWL (DL) rather than a generic approach. Auer and Herre [5] propose a generic framework to support the versioning and evolution of RDF graphs. The main concept introduced in this work are *atomic graphs*, which provides a practical approach for dealing with blank nodes in change sets. Additionally they introduce a formal hierarchical system for structuring a set of changes and evolution patterns leading to the changes of a knowledge base. Cassidy and Ballantine [8] present a version control system for RDF graphs based on the model of the DSCM *Darcs*. Their approach covers the versioning operations *commute*, *revert* and *merge*. In contrast to other DSCM systems the merge operation is implemented using patch commutation, which requires history rewriting and thus loses the the context of the original changes.

[10] `https://www.w3.org/2007/03/layerCake.svg`

TailR as presented in [21] is a system for preserving the history of arbitrary linked data sets. It follows a combined delta and snapshot storage approach. The system is comparable to the approach presented by Frommhold et al. [14], as both system are linear change tracking systems. None of the systems provides support for branches to allow independent evolution of RDF graphs. Graube et al. [15] propose the R43ples approach, which uses named graphs for storing revisions as deltas. For querying and updating the triple store an extended SPARQL protocol language is introduced. R&Wbase [26] is a tool for versioning an RDF graph. It is tracking changes which are stored in individual named graphs which are combined on query time, this makes it impossible to use the system to manage RDF datasets with multiple named graphs. The system also implements an understanding of coexisting branches within a versioning graph, which is very close to the concept of Git. In the *dat*[11] project a tool for distributing and synchronizing data is developed. The aim is mainly on synchronizing any file type peer to peer. It has no support for managing branches and merging diverged versions and is not focusing on RDF data.

Table 1 provides an overview of the related work and compares the presented approaches with regard to their used storage system, quad and blank node support, as well as the possibilities to create multiple branches, merge branches and create distributed setups for collaboration using push and pull mechanisms.

Git4Voc, as proposed by Halilaj et al. [17] is a collection of best practices and tools to support authors in the collaborative creation process of RDF and

[11] `http://dat-data.com/`

| Approach | storage | quad support | bnodes | branches | merge | push/pull |
|---|---|---|---|---|---|---|
| Meinhardt et al. [21] | hybrid | no[a] | yes | no[f] | no | (yes)[h] |
| Frommhold et al. [14] | delta | yes | yes | no[f] | no | no |
| Graube et al. [15] | delta | no[b,c] | (yes)[d] | yes | no | no |
| Vander Sande et al. [26] | delta | no[c] | (yes)[e] | yes | (yes)[g] | no |
| *dat* | chunks | n/a | n/a | no | no | yes |

[a] The granularity of versioning are repositories; [b] Only single graphs are put under version control; [c] The context is used to encode revisions; [d] Blank nodes are skolemized; [e] Blank nodes are addressed by internal identifiers; [f] Only linear change tracking is supported; [g] Naive merge implementation; [h] No pull requests but history replication via memento API

**Table 1.** Comparison of the different (D)SCM systems for RDF data. All of these systems are coming with custom implementations and are not reusing existing SCMs. At the level of abstraction all of these systems can be located on the *data interchange* layer.

OWL vocabularies. The methodology is based on Git and implements pre- and post-commit hooks, which call various tools for automatically checking the vocabulary specification. In addition to the methodology it has also formulated very important requirements for the collaboration on RDF data. Fernández et al. [10] put their focus on the creation of a benchmark for RDF Archives – BEAR. The authors are focusing on *"exploiting the blueprints to build a customizable generator of evolving synthetic RDF data"*. We hope that this can be used as generic benchmark in the future.

## 3 Preliminaries

In the following we are speaking of *RDF graph* resp. just *graph* and *RDF dataset* resp. just *dataset* as commonly used in RDF and also defined in [9].

According to [5] an *Atomic Graph* is defined as follows:

**Definition 1 (Atomic Graph).** *A graph is atomic if it may not be split into two nonempty graphs whose blank nodes are disjoint.*

This means that all graphs containing exactly one statement are atomic. Furthermore the graph is still atomic if it contains a statement with a blank node and all other statements in the same graph also contain this blank node. If a statement as well contains a second blank node the same takes effect for this blank node recursively. Basically an *Atomic Graph* is comparable to a *Minimum Self-Contained Graph* (MSG) as introduced by [29].

We define $\mathbb{G}_A$ as the set of all *Atomic Graphs*. We also define the equivalence relation on two graphs $G, H \in \mathbb{G}_A$ as $G \approx H \Leftrightarrow G$ and $H$ are *isomorphic* as defined for RDF in [9]. Further the quotient set of $\mathbb{G}_A$ by $\approx$ is $\mathbb{A} := \mathbb{G}_A/\approx$, the quotient set of all equivalence classes of atomic graphs. Based on this we now define an *Atomic Partition* of a graph as follows:

**Definition 2 (Atomic Partition).** *Let $\mathcal{P}_G \subseteq \mathbb{G}_A$ denote the partition of $G$ containing only atomic graphs.*

*Then the* Atomic Partition *is defined as*

$$P(G) := \mathcal{P}_G / \approx$$

This means that a graph is split into a set of sets, called partition, containing only atomic graphs. Further we can also say, that $P(G) \subseteq \mathbb{A}$. Each of the containing sets consists of exactly one statement for all statements without blank nodes. For statements with blank nodes, it consists of the complete subgraph connected to a blank node and all neighboring blank nodes. This especially means, that all sets in the *Atomic Partition* are disjoint regarding the contained blank nodes. Further they are disjoint regarding the contained triples (because its a partition).

We now also let $\breve{P}(X)$ be a subset of $X$ consisting of exactly one element of each equivalence class, where $X$ is an *Atomic Partition*. This means that $G$ and $\breve{P}(P(G))$ are isomorphic, we just make sure that no multiple isomorphic atomic subgraphs exist.

**Definition 3 (Difference).** *Let $G$ and $G'$ be two graphs, and $P(G)$ resp. $P(G')$ the* Atomic Partitions.[12]

$$\begin{aligned} C^+ &:= \dot{\bigcup} \left( \breve{P}\left(P(G') \setminus P(G)\right) \right) \\ C^- &:= \dot{\bigcup} \left( \breve{P}\left(P(G) \setminus P(G')\right) \right) \\ \Delta(G, G') &:= (C^+, C^-) \end{aligned}$$

Looking at the resulting tuple $(C^+, C^-)$ we can also say that the inverse of $\Delta(G, G')$ is $\Delta^{-1}(G, G') = \Delta(G', G)$ by swapping the positive and negative sets.

**Definition 4 (Change).** *A change is a tuple of two graphs $(C_G^+, C_G^-)$ on a graph $G$, with*

$$\begin{aligned} &P(C_G^+) \cap P(G) = \emptyset \\ &P(C_G^-) \subseteq P(G) \\ &P(C_G^+) \cap P(C_G^-) = \emptyset \\ &P(C_G^+) \cup P(C_G^-) \neq \emptyset \end{aligned}$$

Since blank nodes can't be identified across graphs, a new blank node has to be completely enclosed in the set of additions. Thus $P(C_G^+)$ and $P(G)$ have to be disjoint. This means an addition can't introduce just new statements to an existing blank node. Parallel to the addition a blank node can only be removed if it is completely removed with all its statements. This is made sure by $P(C_G^-)$ being a subset of $P(G)$. Simple statements without blank nodes can be simply added and removed. Further since $C_G^+$ and $C_G^-$ are disjoint we avoid the removal of atomic graphs, which are added in the same change and vice versa. And since at least one of $C_G^+$ or $C_G^-$ can't be empty we avoid changes with no effect.

[12] The set-minus operator $\setminus$ in this case is defined to also remove elements equivalent with respect to the $\approx$ relation.

**Definition 5 (Application of a Change).** *Let $C_G = (C_G^+, C_G^-)$ be a change on $G$. The function Apl is defined for the arguments $G, C_G$ resp. $G, (C_G^+, C_G^-)$ and is determined by*

$$Apl(G, (C_G^+, C_G^-)) := \dot{\bigcup} \left( \breve{P} \left( (P(G) \setminus P(C_G^-)) \cup P(C_G^+) \right) \right)$$

*We say that $C_G$ is applied to $G$ with the result $G'$.*

## 4 Operations

Now that we have defined the calculation with additions and removals on a graph we can define basic version tracking and distributed evolution operations. Figure 1 depicts an initial commit $\mathcal{A}$ without any predecessor resp. parent commit and a commit $\mathcal{B}$ referring to its parent $\mathcal{A}$.

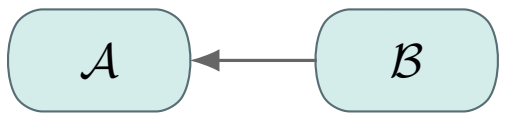

**Figure 1.** Two commits with an ancestor reference

Let $G^0$ be a graph under version control. $\mathcal{A}(\{G^0\})$ is a commit containing the graph $G^0$. $G$ will be the new version of $G^0$ after a change $(C_{G^0}^+, C_{G^0}^-)$ was applied on $G^0$; $Apl(G^0, (C_{G^0}^+, C_{G^0}^-)) = G$. Now we create a new commit containing $G$ which refers to its predecessor commit, from which it was derived: $\mathcal{B}_{\{\mathcal{A}\}}(\{G\})$. Another change applied on $G$ would result in $G'$ and thus a new commit $\mathcal{C}_{\{\mathcal{B}_{\{\mathcal{A}\}}\}}(\{G'\})$ is created. (In the further writing, the indices and arguments of commits are sometimes omitted for better readability, while clarity should still be maintained by using distinct letters.)

The evolution of a *graph* is the process of subsequently applying changes to the graph using the *Apl* function as defined in definition 5. Each commit is expressing the complete evolution process of a set of graphs, since it refers to its predecessor, which in turn refers to its predecessor as well. Initial commits holding the initial version of the graph are not referring to any predecessor.

Since a commit is referring to its predecessor and not vice versa, nothing hinders us from creating another commit $\mathcal{D}_{\{\mathcal{B}_{\{\mathcal{A}\}}\}}(\{G''\})$. Taking the commits $\mathcal{A}$, $\mathcal{B}_{\{\mathcal{A}\}}$, $\mathcal{C}_{\{\mathcal{B}\}}$, and $\mathcal{D}_{\{\mathcal{B}\}}$ results in a directed rooted in-tree, as depicted in fig. 2. The commit $\mathcal{D}$ is now a new *branch* or *fork* based on $\mathcal{B}$, which is diverged from $\mathcal{C}$. We know that $G \neq G'$ and $G \neq G''$, while we don't know about the relation between $G'$ and $G''$. From now on the graph $G$ is *independently* evolving in two branches, while *independent* means possibly independent, i. e. two actors performing a change do not have to know of each other or do not need a direct communication channel. The actors could actually communicate, but communication is not required for those actions.

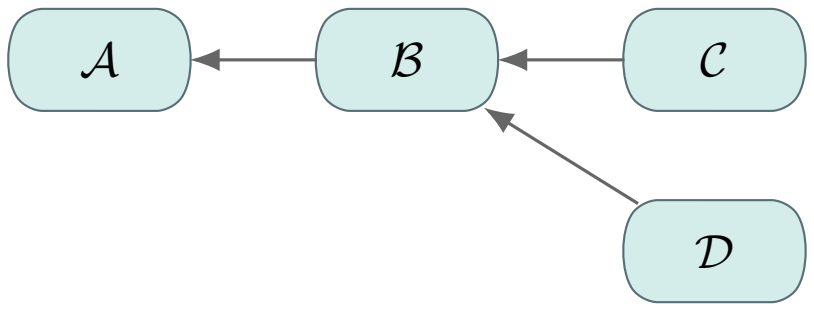

**Figure 2.** Two branches evolved from a common commit

**Definition 6 (Distributed Evolution).** *Distributed Evolution is the (independent) evolution of a* graph $G$ *with two graphs* $G_1$ *and* $G_2$ *as result, where* $Apl(G, C_1) = G_1$ *and* $Apl(G, C_2) = G_2$*. The changes* $C_1$ *and* $C_2$ *might be unequal, but can be the same. The same applies for* $G_1$ *and* $G_2$*, they can be different after the independent evolution, but can be similar.*

### 4.1 Merge Different Branches

After creating a second branch, the tree of commits is diverged, as shown in the example of fig. 2. We now want to merge the branches again, in order to get a version of the graph, containing changes made in those different branches or at least take all of these changes into account. The notation of the merge is defined as follows:

**Definition 7 (Merge of two Evolved Graphs).** *Given are two commits* $\mathbb{B}_{\{\mathbb{X}\}}(\{G\})$ *and* $\mathbb{D}_{\{\mathbb{C}\}}(\{H\})$ *Merging the two* graphs $G$ *and* $H$ *with respect to the change history expressed by the commits* $\mathbb{B}$ *and* $\mathbb{D}$ *is a function*

$$Merge(\mathbb{B}(\{G\}), \mathbb{D}(\{H\})) = \mathbb{E}_{\{\mathbb{B},\mathbb{D}\}}(\{I\})$$

*With* $\mathbb{E}_{\{\mathbb{B},\mathbb{D}\}}(\{I\})$ *being the merge commit and* $I$ *the merged graph resulting from* $G$ *and* $H$*.*

The $Merge$ function is taking two commits as arguments and creates a new commit dependent on the input commits. If we take our running example this is done by creating a commit $\mathcal{E}_{\{\mathcal{C}_{\{\mathcal{B}\}},\mathcal{D}_{\{\mathcal{B}\}}\}}(\{G'''\})$, which has two predecessor commits it is referring to. Taking the commits $\mathcal{A}$, $\mathcal{B}_{\{\mathcal{A}\}}$, $\mathcal{C}_{\{\mathcal{B}\}}$, $\mathcal{D}_{\{\mathcal{B}\}}$, and $\mathcal{E}_{\{\mathcal{C},\mathcal{D}\}}$, we get an acyclic directed graph, as it is depicted in fig. 3.

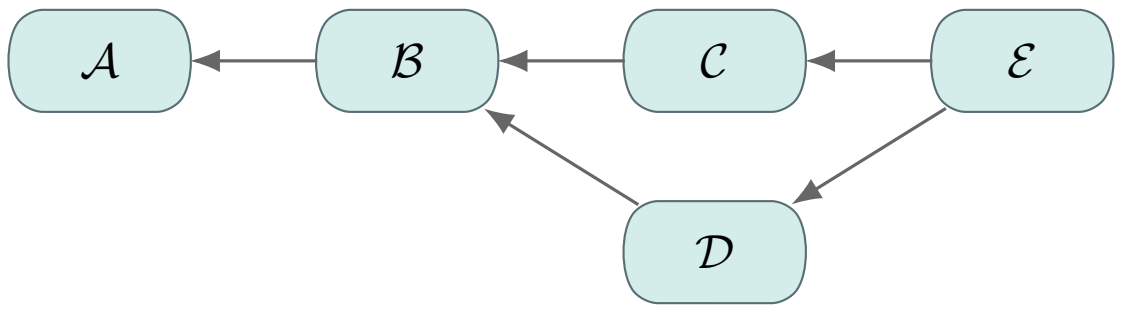

**Figure 3.** Merging commits from two branches into a common version of the graph

Note that the definition doesn't make any assumptions about the predecessors of the two input commits. It depends on the actual implementation of the $Merge$ function whether it is required that both commits have any common ancestors. Furthermore different merge strategies can produce different results, thus it is possible to have multiple merge commits with different resulting graphs but with the same ancestors. Possible merge strategies are presented in section 5.

### 4.2 Revert a Commit

Reverting the commit $\mathcal{B}_{\{\mathcal{A}\}}(\{G\})$ is done by creating an inverse commit $\mathcal{B}^{-1}_{\{\mathcal{B}\}}(\{\tilde{G}^0\})$ (while the commit $\mathcal{A}$ is specified as $\mathcal{A}(\{G^0\})$). This inverse commit is then directly applied to $\mathcal{B}$. The resulting graph $\tilde{G}^0$ is calculated by taking the inverse difference $\Delta^{-1}(G^0, G) = \Delta(G, G^0)$ and applying the resulting change to $G$. After this operation $\tilde{G}^0 = G^0$.

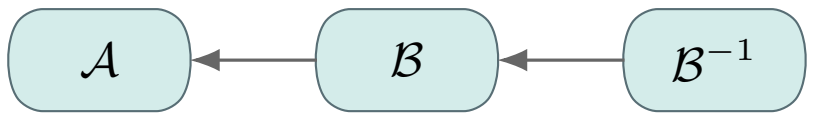

**Figure 4.** A commit reverting the previous commit

A versioning log containing three commits is shown in fig. 4. The latest commit reverts its parent and thus the state of the working directory is again equal to the one expressed by the first commit. While it is obvious how to revert the previous commits, it might be a problem if other commits exist between the commit to be reverted and the current top of the versioning log. In this case a three-way-merge is applied (cf. section 5). For this merge, the merge *base* is the commit to be reverted, branch $\mathcal{A}$ is the parent commit of the commit which is to be reverted, and branch $\mathcal{B}$ the currently latest commit.

## 5 Merge Strategies

Since we are not only interested in the abstract branching and merging model of the commits, we want to know what a merge operation means for the created graph in the commit. In the following we present some possible implementations of the merge operations. Note, that *merging* in this context is not–generally–to be understood as in the *RDF 1.1 Semantics Recommendation* [18] as the union of two graphs.

**Union Merge** Merging two *graphs* $G'$ and $G''$ could be considered trivially as the union operations of the two graphs: $G' \cup G'' = G'''$. This merge–as mentioned above–is well defined in the RDF 1.1 Semantics Recommendation [18] in section "4.1 Shared blank nodes, unions and merges". But this operation would not take into account the actual change operations leading to the versions of the graphs. Furthermore it does not allow the implementation of conflict detection or resolution operations, which might be intended in unsupervised situations.

**All Ours/All Theirs** Two other merge strategies, which wouldn't produce merge conflicts are *ours* and *theirs*, which just take the complete graph $G' = G'''$ or $G'' = G'''$, while ignoring the other graph respectively. This strategy might be chosen to completely discard changes from a certain branch.

**Three-Way Merge** A methodology used in distributed version control systems for software source code, such as Git and Mercurial is the *three-way-merge*[13]. The merge consists of three phases, (1) finding a common *merge base* for the two commits to merge, (2) inferring which lines where added and removed between the *merge base* and the individual branches, and (3) creating a merged version by combining the changes made in the two branches.

Taking into account the versions of the graphs in the two commits to be merged $A_{\{D\}}(\{G'\})$, $B_{\{E\}}(\{G''\})$ we find the most recent common ancestor $F(\{G\})$. This strategy relies on the existence of a common ancestor $F$, such that for $A$ and $B$ there must exist an ancestor path $A\left\{\ddots_{\{F,\dots\}}\right\}$ resp. $B\left\{\ddots_{\{F,\dots\}}\right\}$ to $F$.

$$
\begin{aligned}
(C_A^+, C_A^-) &= \Delta(G, G') \\
(C_B^+, C_B^-) &= \Delta(G, G'') \\
G''' &= \bigcup \left( (P(G') \cap P(G'')) \cup P(C_A^+) \cup P(C_B^+) \right)
\end{aligned}
$$

With this merge strategy in Git changes from two different branches on the same line or two neighboring lines are marked as conflicts. In this case Git can't decide how to combine the changes or which line to take. For transferring this principle from source code files to graphs, we first see, that this definition of conflict is related to its *local* context. In graphs the local context for a statement can be other outgoing and incoming edges of its subject and object resources.

[13] How does Git merge work: `https://www.quora.com/How-does-Git-merge-work`, 2016-05-10

Usually in source code management with this merge strategy a merge conflict occurs, when two close by lines were added or removed in different branches. Thus it can't be decided, whether they are resulting from the same original line or if they are even contradicting. This means for two close by inserted lines one can't automatically decide on in which order the lines have to be inserted into the merge result. Since there is no order relevant in the RDF data model, we don't have to produce merge conflicts in those situations. Producing merge conflicts in a graph's local context is subject to future work.

**Touch Merge** This merge strategy is an extension of the *Three-Way Merge*, which should introduce an additional manual revision of the merge operation. In a situation, where we have the commits $A(\{G\})$, $B_{\{A\}}(\{G'\})$, $C_{\{B\}}(\{G''\})$, $D_{\{A\}}(\{G'''\})$ as depicted in fig. 5. We can have an atomic subgraph $g$[14], which was not in the merge base (common ancestor, $A$), but was added in both commits $B$ and $D$, while it was again removed in $C$. (Following $g \nsubseteq G \vee G''$, $g \subseteq G' \wedge G'''$.) When merging $C$ and $D$ the *Three-Way Merge* would decide on including $g$ in the result, since it would assume it as added in $D$ but doesn't see it in $C$. The intention of the editors might have been, that $g$ was true at the point in time, when $B$ and $D$ were created, but the creator of $C$ was already updated, that $g$ doesn't hold anymore. The *Touch Merge* (as presented in algorithm 1) in addition to dealing with adds and removals, also marks atomic graphs as *touched*, once they were changed in a commit and presents them as conflict in case of doubt.

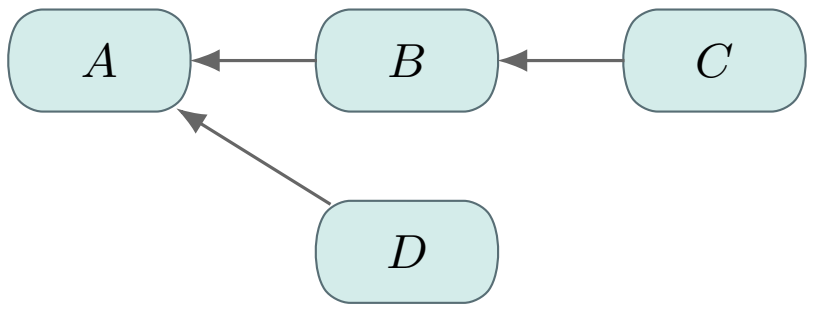

**Figure 5.** A situation of one commit revising a priorly introduced change

Taking into account the two commits to be merged $A_{\{D\}}(\{G'\})$, $B_{\{E\}}(\{G''\})$ we find the most recent common ancestor $F(\{G\})$. Now we run algorithm 1 with $A_{\{D\}}(\{G'\})$ and $F(\{G\})$ and store the results as follows $+_A := +, -_A := -$; we do the same for $B_{\{E\}}(\{G''\})$ and $F(\{G\})$ and store the results as $+_B := +, -_B := -$. The result of the merge will be the graph $G'''$.

$$G''' = \bigcup ((P(G') \cap P(G'')) \cup (+_A \setminus -_B) \cup (+_B \setminus -_A))$$

The conflicting changes, which have to be manually resolved are in a tuple $(Y_1, Y_2)$. $Y_1$ are the atomic graphs added in $A$, but removed in $B$ and $Y_2$ are the atomic graphs added in $B$ but removed in $A$.

[14] The subgraph is denoted by a lowercase letter to underline, that it represents an atomic unit of change.

**Data:** $A$, $F$
**Result:** $+$, $-$
Let $X$ be a sequence of all commits $x_i \in X; 0 \leq i \leq n, n > 0$ on the path from $A$ to $F$, ordered from $x_0 := A$ to $x_n := F$;
$+, - = \{\}$ // initialize $+$ and $-$ as two empty sets;
$i = n$;
**while** $i > 0$ **do**
  $G :=$ the graph in $x_{i-1}$;
  $G' :=$ the graph in $x_i$;
  $(C^+, C^-) := \Delta(G, G')$;
  $+ = (+ \setminus P(C^-)) \cup P(C^+)$;
  $- = (- \setminus P(C^+)) \cup P(C^-)$;
  $i = i - 1$;
**end**

**Algorithm 1:** The touch algorithm for a path from commit $A$ to commit $F$

$$(Y_1, Y_2) = (+_A \cap -_B, +_B \cap -_A)$$

## 6 Evaluation

For evaluating the proposed framework we consider the correctness of the framework regarding the recorded changes and the performance, memory and storage footprint. In order to pursue this task we have taken our implementation of the *Quit Store* [3]. This currently is a prototypical implementation to prove the concept of our framework, thus we are not aiming at competitive performance results. The implementation is written in Python[15]. It is based on the RDFlib[16] for handling RDF, performing the query processing and also uses its in-memory quad store implementation. The HTTP interface to provide a *SPARQL 1.1 Endpoint* is using the Flask API[17].

The hardware setup of the machine running the benchmarks is a virtual machine on a Hyper-V cluster with Intel(R) Xeon(R) CPU E5-2650 v3 with a maximum frequency of 2.30$GHz$ and 62.9$GiB$ of main memory. As operating system Ubuntu 16.10 (yakkety) 64-Bit is used.

```
$ ./generate -pc 4000 -ud -tc 4000 -ppt 1
```

**Listing 1.1.** The BSBM generate command with its argument

As benchmarking framework we have decided to use the Berlin SPARQL benchmark (BSBM) [7], since it is made for executing SPARQL Query and SPARQL Update operations. The initial dataset as it is generated using the

[15] `https://www.python.org/`
[16] `https://rdflib.readthedocs.io/en/stable/`
[17] `http://flask.pocoo.org/`

BSBM, shown in listing 1.1, contains 46370 statements and 1379201 statements to be added and removed during the benchmark. To also execute update operations we are using the *Explore and Update Use Case*. We have executed 40 warm-up and 1500 query mix runs which resulted in 4592 commits on the underlying git repository using the testdriver as shown in listing 1.2.

```
$ ./testdriver http://localhost:5000/sparql \
  -runs 1500 -w 40 -dg "urn:bsbm" -o run.xml \
  -ucf usecases/exploreAndUpdate/sparql.txt \
  -udataset dataset_update.nt \
  -u http://localhost:5000/sparql
```

**Listing 1.2.** The BSBM testdriver command with its argument

The setup for reproducing the evaluation is also available at the following link: `https://github.com/AKSW/QuitEval`.

### 6.1 Correctness of Version Tracking

For checking the correctness of the recorded changes in the underlying git repository we have created a verification setup. The verification setup takes the git repository, the initial data set and the query execution log (`run.log`) produced by the BSBM setup. The repository is set to its initial commit, while the reference store is initialized with the initial dataset. Each update query in the execution log is applied to the reference store. When an effect, change in number of statements, is detected on the store, the git repository is forwarded to the next commit. Now the content of the reference store is serialized and compared to the content of the git repository at this point in time. This scenario is implemented in the `verify.py` script in the evaluation tool collection. We have executed this verification scenario and could verify, that the recorded repository has the same data as the store after executing the same queries.

### 6.2 Performance

The performance of the reference implementation was analyzed in order to identify obstacles in the conception of our approach. In fig. 6 the queries per second for the different categories of queries in the BSBM are given and compared to the baseline. We compare the execution of our store with version tracking enabled to only the in memory store of python rdflib without version tracking. As expected the versioning has a big impact on the update queries (INSERT DATA and DELETE WHERE), while the explore queries (SELECT, CONSTRUCT and DESCRIBE) are not further impacted. We could reach $83QMpH$[18] resp. $82QMpH$ with quit versioning (without resp. with garbage collection) and $787QMpH$ for the baseline.

[18] $QMpH$ Query Mixes per Hour, Query Mixes are defined by the BSBM

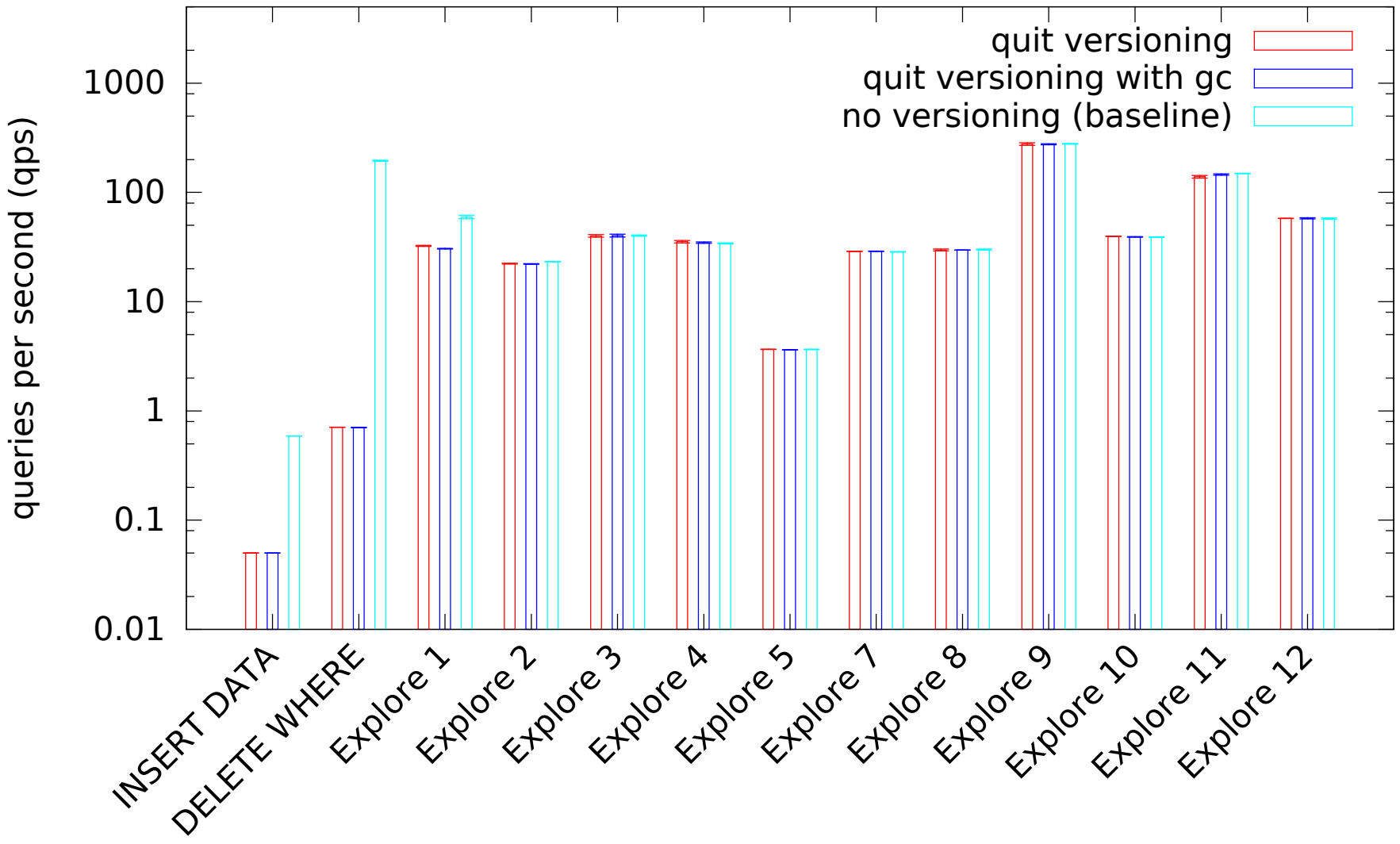

**Figure 6.** Execution of the different BSBM queries.

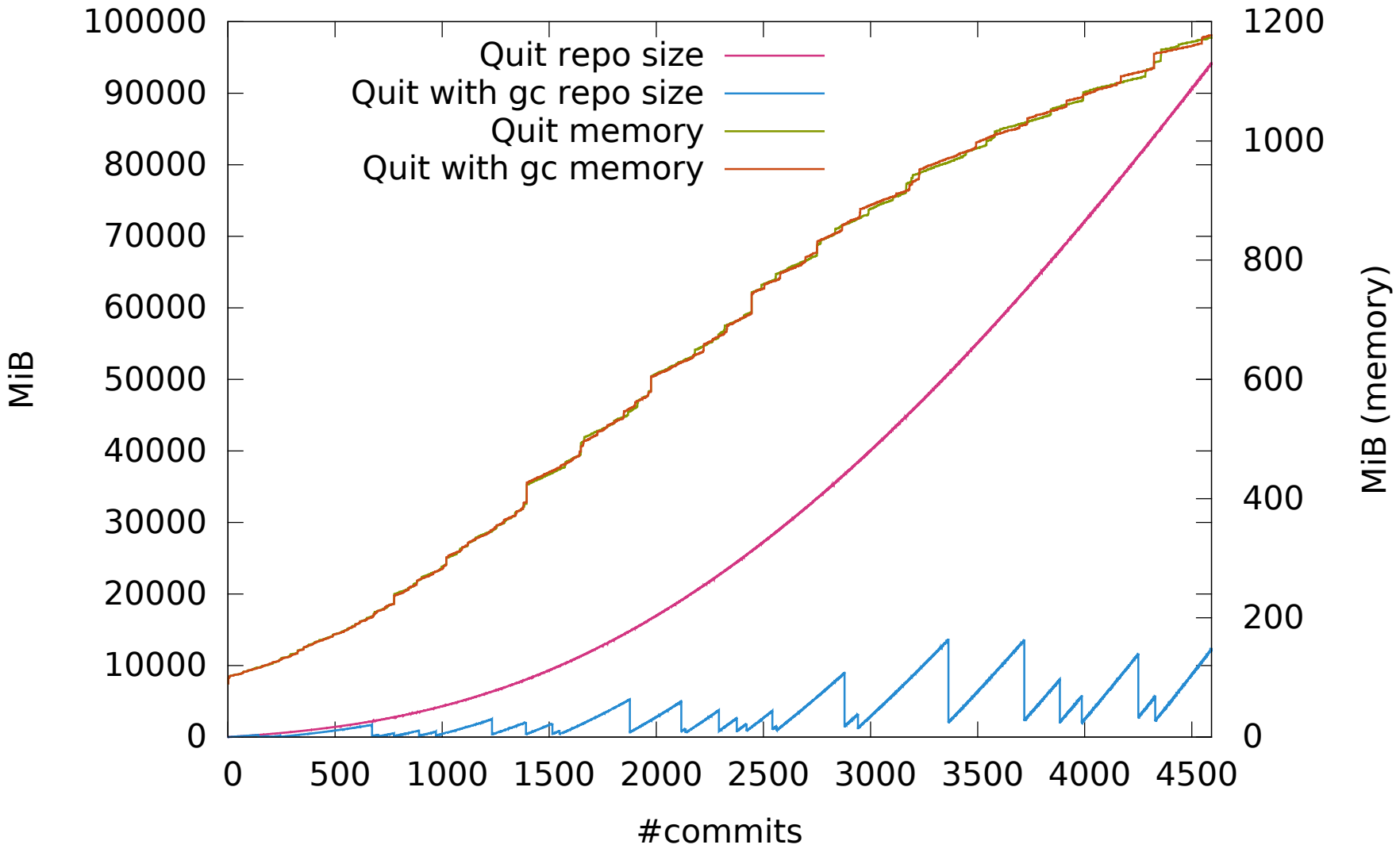

**Figure 7.** Usage of system resources and number of commits in the repository during the execution of the BSBM. (gc = garbage collection)

Additionally we have monitored the impact of the execution of the store on the system. The results are visualized in fig. 7. We have measured the size of the repository, the memory usage of the store process during the execution of the BSBM and have put it in relation to the number of commits generated at that point in time. The memory consumption increased from $100MiB$ to $1.1GiB$ during the execution. The repository has increased from initially $15MiB$ to a size of $92GiB$ without garbage collection at the end of the benchmark. Enabling the garbage collection on the quit store during the benchmark could compress the repository to $12GiB$.

### 6.3 Correctness of the Merge Method

The functional correctness of the merge method was verified using a repository filled with data using the graph generated by the BSBM. To create the evaluation setup and run the verification of the results, a script was created. This script takes a git repository and creates a setup of three commits. An initial commit contains a graph file, which serves as base of two branches. Each of the branches is forked from this initial commit and contains an altered graph file. The files in the individual commits contain random combinations of added and removed statements, while the script also produces the graph, which is expected after merging the branches.

After creating the two branches with different graphs they are merged using `git merge`. The result is then compared to the expected graph and the result is presented to the user. We have executed the verification 1000 times and no merge conflict or failure in the merge result occurred.

## 7 Conclusion and Future Work

In this paper we have presented a formal framework for the distributed evolution of RDF knowledge bases. It provides a definition of atomic operations for applying on RDF graphs as well as formalized definitions of the versioning operations *commit*, *branch*, *merge* and *revert*. In contrast to the related work in section 2 our approach is quad aware, can handle blank nodes, supports branches, supports merging with conflict resolution and allows distributed collaboration with push and pull operations. Also the naive merge strategies *Union*, *All Ours* and *All Theirs* are described. The *Three-Way Merge*, as it is used in Git is then transfered to the application on atomic graphs to be used on RDF datasets. For allowing more manual control over the merge process also the *Touch Merge* was introduced. The provided formal model and methodology is implemented in the Quit Store tool, which provides a SPARQL 1.1 read/write interface to query and update an RDF dataset in a quad store. The contents of the store are tracked for version control in Git in the local filesystem in parallel. Based on this prototypical implementation we have pursued an evaluation regarding the correctness of the model and monitored the performance of our implementation. The theoretical foundations could be confirmed, while the results regarding the performance reveal potential for improvement. Using the garbage collection we could show that the snap shot based approach doesn't put high load on storage requirements.

Based on the presented approach the application in distributed and collaborative data curation scenarios is now possible. It enables the setup of platforms similar to GitHub specialized on the needs of data scientists and data engineers for creating datasets using local working copies and sending pull requests, while the versioning history and provenance of the results is automatically tracked. Also this system can support the application of RDF in enterprise scenarios such as supply chain management as described in [13]. An integration with the distributed evolution model of distributed semantic social network [4,27], as well as the use-case of synchronization in mobile scenarios [28] is possible. Further we are planning to lift the collaborative curation and annotation in distributed scenarios such as presented in the Structured Feedback protocol [1] to the next level by directly recording the user feedback as commits, which can enable powerfull co-evolution strategies.

## 8 Acknowledgement

We want to thank Sören Auer for his valuable important remarks and Norman Radtke for his precious and tireless implementation work. This work was partly supported by a grant from the German Federal Ministry of Education and Research (BMBF) for the LEDS Project under grant agreement No 03WKCG11C.